\def\et{{et al.\ }}
\def\mcg{{MCG--6-30-15}}

\def\asca{{\it ASCA}}
\def\xmm{{\it XMM-Newton}}

\def\Msun{\hbox{$\rm\thinspace M_{\odot}$}}

\documentclass{kluwer}    

\newdisplay{guess}{Conjecture}
\usepackage{psfig}

\begin{document}                                                                                   
\begin{article}
\begin{opening}         
\title{Broad Iron Lines in AGN and X-ray Binaries} 
\author{A. C.  \surname{Fabian}}  
\runningauthor{A. C. Fabian}
\runningtitle{Broad Iron Lines}
\institute{Institute of Astronomy, Madingley Road Cambridge CB3 0HA, UK}

\begin{abstract}

Several AGN and black hole X-ray binaries show a clear very broad iron
line which is strong evidence that the black holes are rapidly
spinning. Detailed analysis of these objects shows that the emission
line is not significantly affected by absorption and that the source 
variability is principally due to variation in amplitude of a
power-law. Underlying this is a much less variable,
relativistically-smeared, reflection-dominated, component which
carries the imprint of strong gravity at a few gravitational radii.
The strong gravitational light bending in these regions then explains
the power-law variability as due to changes in height of the primary
X-ray source above the disc.  The reflection component, in particular
its variability and the profile of the iron line, enables us to 
study the innermost regions around an accreting, spinning, black
hole. 

\end{abstract}

\end{opening}           

\section{Introduction}  

Radiatively-efficient accreting black holes are expected to be
surrounded by a dense disc radiating quasi-blackbody thermal EUV and
soft X-ray emission. Hard X-ray emission originates via Comptonization
of that soft radiation in a corona above the disc, fed by magnetic
fields from the body of the disc. Irradiation of the dense disc
material by hard X-rays then gives rise to a characteristic
`reflection' spectrum, computed examples of which are shown in Fig.~1
(from Ross \& Fabian 2004). 

Most of the power is radiated from close to the smallest disc radii
which for a non-spinning black hole is $6 r_{\rm g}$, where $r_{\rm
g}=GM/c^2$. For a spinning (Kerr) black hole it reduces as the spin
increases (Bardeen et al. 1972) to 1.23 $r_{\rm g}$ for what is assume
to be maximal spin (Thorne 1972). Relativistic effects then affect the
appearance of the reflection spectrum through Doppler, aberration,
gravitational redshift and light bending effects (Fabian et al. 1989;
Laor 1991). The dominant feature in the spectrum seen by a distant
observer is an iron line with a broad skewed profile.

Broad iron lines seen in the spectrum of several active galaxies and
Galactic black hole binaries are reviewed here. The cases for
relativistic lines in the Seyfert galaxy MCG--6-30-15 and the X-ray
binary GX\,339-4 are very strong, indicating that those black holes
are rapidly spinning. The puzzling spectral variability of such
sources is now beginning to be understood within the context of
emission from the strong gravity regime (Miniutti \& Fabian 2004).
Some active galactic nuclei (AGN) and X-ray black hole binaries show
either no line or only a narrow one.  This is discussed within the
context of state changes and jetted emission observed in the Galactic
black holes.

\begin{figure}[t] 
\begin{center} 
\hbox{
\psfig{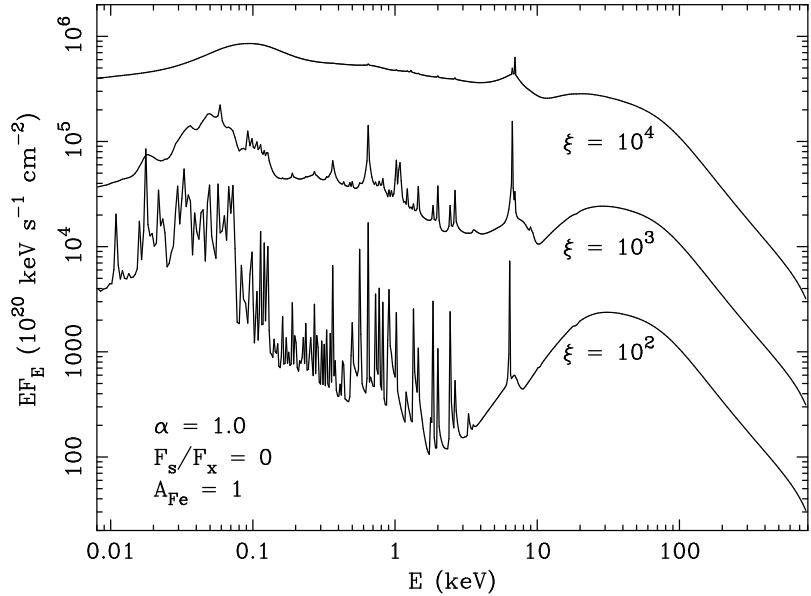}
}
\vspace{-1.0cm} 
\end{center} 
\caption{\footnotesize {Computed reflection spectrum as a function of
$\xi=F/n,$ where $n$ is the density of the surface (Ross \& Fabian
2004).} }
\label{fig1} 
\end{figure}

\section{MCG--6-30-15}

The X-ray spectrum of the bright Seyfert 1 galaxy \mcg\ ($z=0.00775$)
has a broad emission feature stretching from below 4~keV to about
7~keV. The shape of this feature, first clearly resolved with \asca\
by Tanaka \et (1995), is skewed and peaks at about 6.4~keV. This
profile is consistent with that predicted from iron fluorescence from
an accretion disc inclined at 30~deg extending down to within about 6
gravitational radii ($6r_{\rm g} = 6GM/c^2$) of a black hole (Fabian
\et 1989; Laor 1991). In part of the \asca\ observation the line
extended below 4~keV (Iwasawa \et 1996) which means that the emission
originates at radii much less than $6r_{\rm g}$, probably due to the
black hole spinning. \xmm\ has observed \mcg\ twice (in 2000, Wilms
\et 2001 and 2001; Fabian \et 2002a) and in both cases the line
extended down to about 3~keV (Fig.~2a), implying the spin parameter
$a>0.93$ (Dabrowski et al. 1997; Reynolds et al. 2004). This raises the
exciting possibility that the spin energy of the hole is being tapped
(Wilms et al. 2001). 

The X-ray continuum emission of \mcg\ is highly variable (see Vaughan,
Fabian \& Nandra 2003, Vaughan \& Fabian 2004 and Reynolds et al. 2004
for recent analyses). If the observed continuum drives the iron
fluorescence then the line flux should respond to variations in the
incident continuum on timescales comparable to the light-crossing, or
hydrodynamical time of the inner accretion disc (Fabian et al. 1989;
Stella 1990; Matt \& Perola 1992; Reynolds et al. 1999). This timescale
($\sim 100M_6$~s for reflection from within $10r_{\rm g}$ around a
black hole of mass $10^6 M_6$~\Msun) is short enough that a single,
long observation spans many light-crossing times. This has motivated
observational efforts to find variations in the line flux (e.g.
Iwasawa \et 1996, 1999; Reynolds 2000; Vaughan \& Edelson 2001; Shih,
Iwasawa \& Fabian 2002). These analyses indicated that the iron line
in \mcg\ is indeed variable on timescales of $\sim 10^4$~s (e.g.
Fig.~3), but that the general amplitude of the variations was
considerably less than expected and not directly correlated with the
observed continuum. 

\begin{figure}[]
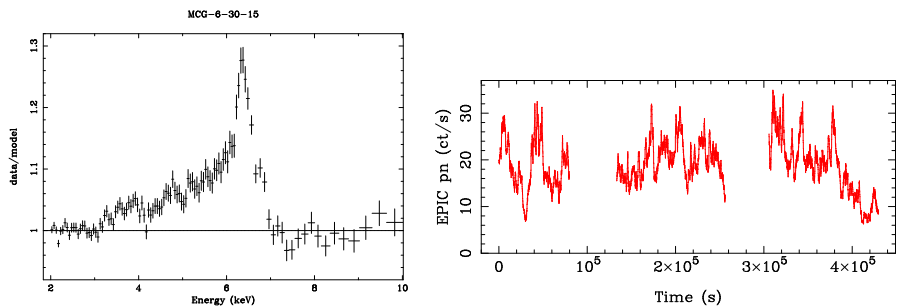
 
\begin{center} 
\hbox{
\psfig{figure=mcg6_linebw.ps,width=0.45\textwidth,angle=-90}
\hspace{0.1cm}
\psfig{figure=lca.ps,width=0.52\textwidth,angle=-90}
\hspace{1.0cm}

} \vspace{-1.0cm} \end{center} \caption{\footnotesize {{\it Left
panel}: The broad iron line in \mcg\ from 2001 (Fabian et al. 2002).
{\it Right panel}: The light curve in 2001.}} \label{fig1} 
\end{figure}

\begin{figure}[]
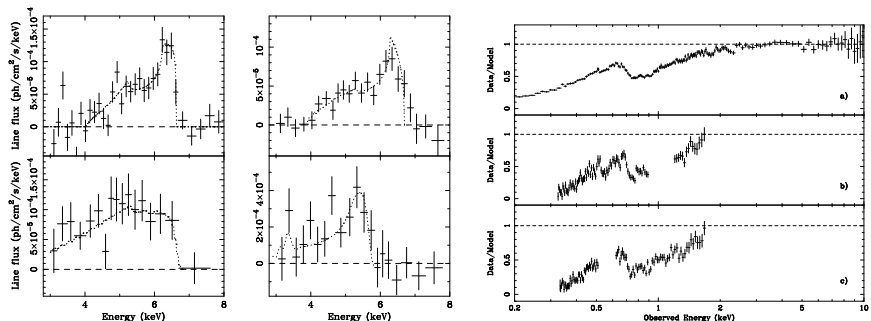

\begin{center}
\hbox{\hspace{0cm}
\psfig{figure=mcg6profs.ps,width=0.5\textwidth,angle=-90} 
\hspace{0.2cm}
\psfig{figure=combined_absorption_2.eps,width=0.43\textwidth,angle=-90}

} 
\vspace{-1.0cm} 
\end{center} 
\caption{\footnotesize 
{{\it Left panel}: Line profile variations in \mcg\ seen with ASCA in
1994 (left) and 1996 (right). {\it Middle panel}: Iron line
variability in the last orbit of 2001 (from Iwasawa et al. 2004). {\it
Right panel} EPIC difference spectrum between the brighter and dimmer
parts of the long XMM observation, presented as a ratio to a power-law
model fitted over the 3--10~keV band, with similar spectra for RGS1
and 2 below (from Turner et al. 2004). } }
\label{fig1} 
\end{figure}

\begin{figure}[t]
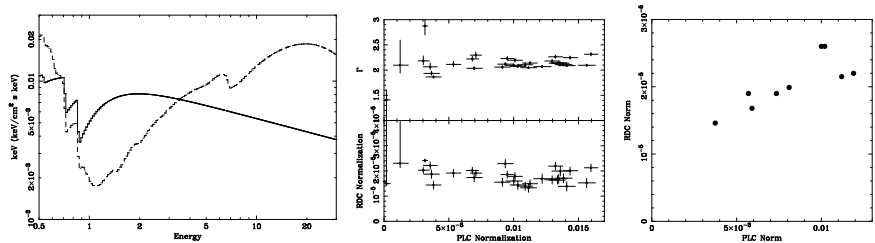

\begin{center}
\hbox{
\psfig{figure=specdecomp.ps,width=0.37\textwidth,angle=-90}
\hspace{0.cm}
\psfig{figure=normplot.ps,width=0.28\textwidth,angle=-90}
\hspace{0.cm}
\psfig{figure=rdcplc.ps,width=0.28\textwidth,angle=-90}
}
\vspace{-1.0cm}
\end{center}
\caption{\footnotesize
{{\it Left panel}: The two component model in which the PLC (solid
line) varies considerably in amplitude while the RDC (dashed line)
varies little.
{\it Centre panel}: top: Photon index $\Gamma$ of the PLC plotted against
its normalization, bottom: RDC normalization plotted against PLC
normalization for 2001.
{\it Right panel}: RDC vs PLC normalizations for 2000. }}
\label{fig1}
\end{figure}

The long \xmm\ observation (Fabian et al. 2002) showed that a simple
two-component model (Shih et al. 2002) is sufficient to explain the
observed spectral variability (Fig.~1b; Fabian \& Vaughan 2003; Taylor
et al. 2003). The model consists of a highly variable power-law
component (PLC) plus a much less variable harder component carrying
the iron line (RDC, Fig.~2a). It gives an excellent fit to the data,
with the harder, line-carrying component dominating lowest flux states
of the observation (Fabian \& Vaughan 2003). That the variation is
driven by a power-law is evident from difference spectra made by
subtracting the spectra of fainter parts of the lightcurve from those
of brighter parts and fitting the resulting 'difference spectrum'. It
is a power-law from 3--10 keV with no iron-K features. On the
assumption that this power-law continues to lower energies, where
attenuation at low energies due to both Galactic absorption and the
warm absorber in \mcg\ is seen. This demonstrates that there is no
subtle additional absorption influencing the shape of the extensive
low-energy ``red'' wing to the iron line. Small variations in the
amplitide of the RDC are seen.

A detailed analysis of the XMM-Newton 2001 data by Turner et al. (2003,
2004) shows, from a curve of growth analysis of the absorption lines
and difference spectra, that the warm absorber accounts for most of the
soft X-ray spectral features and that any distinct relativistically-broadened
CNO lines (Branduardi-Raymont et al. 2001; Sako et al. 2003) are weak.

\subsection{\it Interpretation}

Explaining the relatively small variability of the RDC, compared with
that of the PLC, provides a significant challenge. It appears to be
mostly due to reflection but it is not simple reflection of the
observed power-law component since that repeatedly varies by factors
of two or more on short timescales; the RDC and PLC appear partially
disconnected. Since however both show the effects of the warm absorber
they must originate in a similar location. As the extensive red wing
of the iron line in the RDC indicates emission peaking at only a few
gravitational radii ($GM/c^2$) we assume that this is indeed where
that component originates.  In such extreme gravity the general
relativistic bending of light is very large, boosting the strength of
reflection (Martocchia et al. 2000, 2002; Dabrowski \& Lasenby 2001)
and can account for the behaviour of the components (Fabian \& Vaughan
2003; Miniutti et al. 2003, 2004).  How bright the PLC appears depends
strongly on its height above the disc. Much of the radiation is bent
down to the disc and black hole when the PLC is at a height of a few
$r_{\rm g}$ but less so above $20 r_{\rm g}$ (Fig.~5a).

\begin{figure}[h] \begin{center} \hbox{
\hspace{0cm}
\psfig{figure=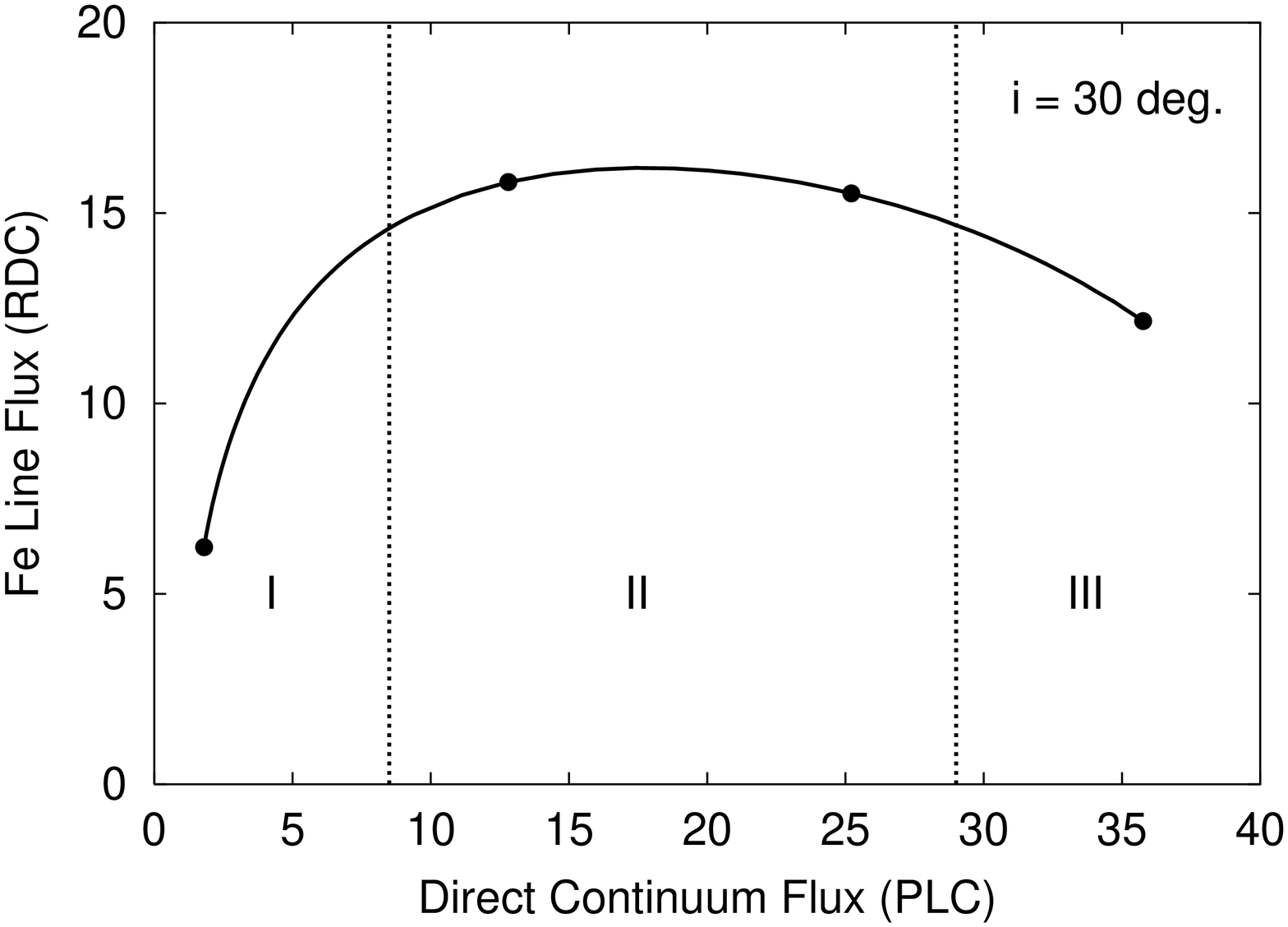,width=0.45\textwidth,angle=-90}
\hspace{0cm}
\psfig{figure=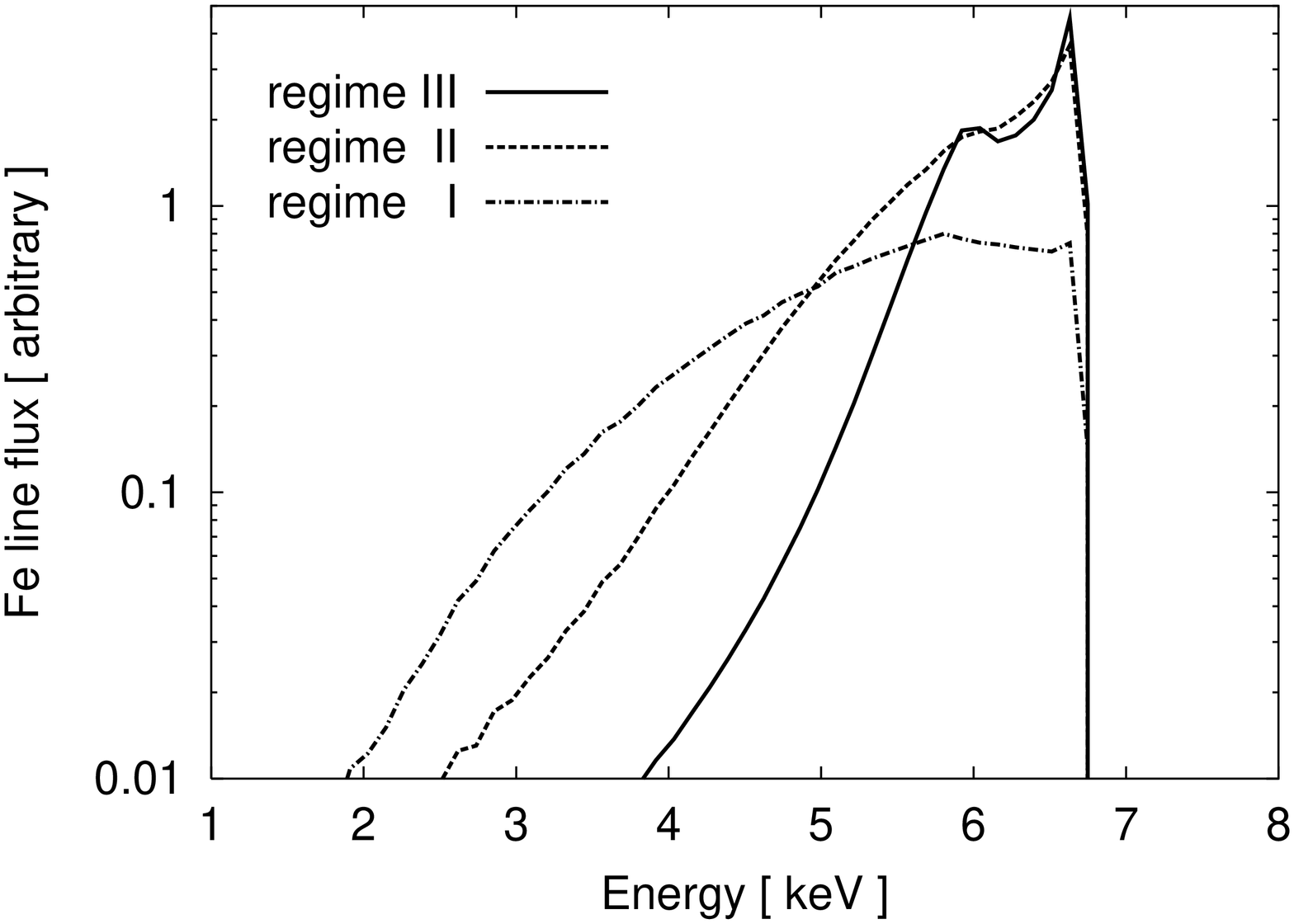,width=0.45\textwidth,angle=-90} }
\vspace{-1.0cm} \end{center} \caption{\footnotesize {{\it Left panel}:
Variation in amplitude of the RDC with height (knots at 1, 5, 10 and
20 gravitational radii) of the PLC. {\it Right panel}: Line profile
changes with PLC height (Miniutti \& Fabian 2004).}} \label{fig1}
\end{figure}

Part of the source variability can thus be explained by an
intrinsically constant PLC changing height above the disc.  Intrinsic
variability of the PLC might also be present. The RDC is expected to
change little during PLC variations due to source position but
will change with intrinsic variability. Line profile changes with
source height is a discriminant (Fig.~5b). 

Tapping of black holes spin by magnetic fields in the disc is a strong
possibility to account for the peaking of the power so close to the
hole (Wilms et al. 2001; Reynolds et al. 2004). 

\section{Galactic Black Hole binaries and NLS1}

Broad iron lines have been found in several Galactic black hole
binaries (or Black Hole Candidates BHC). The lines in GX\,339-4
(Fig.~6a, Miller et al. 2004a,b) and XTE\,J1650-500 (Fig.~6b, Miller et
al. 2002; Miniutti et al. 2004) are among the best examples (see also
Miller et al. 2003, 2002a,b; Martocchia et al. 2002). That in GX\,339-4
shows a very broad red wing indicating that the black hole is rapidly
spinning. Changes in the strength of the iron line as the power-law
continuum varied during the outburst of XTE\,J1650-500 (Rossi et al
2003) follow the sense of the variation expected from the
light-bending model (Fig.~5a). 

\begin{figure}[h]
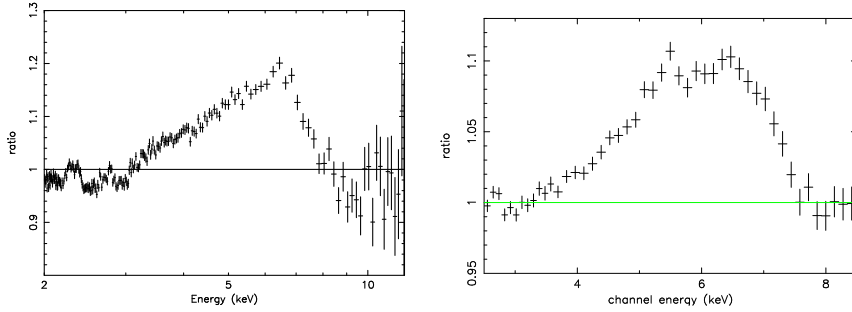
 
\begin{center} \hbox{
\psfig{figure=gx339_xmm.ps,width=0.45\textwidth,angle=-90}
\hspace{0.2cm}
\psfig{figure=line_1_1650.eps,width=0.47\textwidth,angle=-90}
\hspace{1.0cm}
}
\vspace{-1.0cm} 
\end{center} 
\caption{\footnotesize 
{{\it Left panel}: The line in the BHC GX\,339-4 (Miller et al. 2004). 
{\it Right panel}: The broad iron line in XTE\,J1650-500 (Miniutti et
al. 2004).
} }
\label{fig1} 
\end{figure}

\begin{figure}[h]
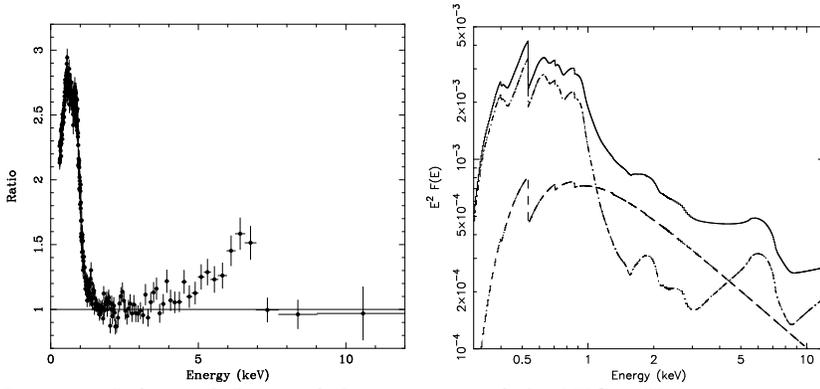
 
\begin{center} \hbox{
\psfig{figure=powrat.ps,width=0.45\textwidth,angle=-90}
\hspace{0.cm}
\psfig{figure=1h0707decomp.ps,width=0.45\textwidth,angle=-90}
\hspace{0cm}

}
\vspace{-1.0cm} 
\end{center} 
\caption{\footnotesize {{\it Left panel}: Ratio of the spectrum of the
NLS1
1H0707 to a power-law fitted between 2 and 3~keV and above 7.5~keV.
{\it Right panel}: Spectral decomposition of 1\,H0707-495 in terms of a
variable power-law (dashed) and a blurred reflection component
(dot-dashed) (from Fabian et al. 2004).
}}
\label{fig1} 
\end{figure}

Narrow Line Seyfert 1 galaxies tend to show steep soft X-ray spectra
and sometimes broad iron emission features. One extreme such object is
1H\,0707-495 which has a marked drop in its spectrum above 7~keV. This
is either an absorption edge showing partial-covering in the source
(Boller et al. 2002, 2004) or the blue wing of a massive, very broad,
iron line (Fig.~7a, Fabian et al. 2002, 2004). A two component,
relativistically-blurred reflection plus power-law, model explains all
the complex spectrum, including its soft excess and broad line,
together with its rapid spectral variability. 1H\,0707-495 is
therefore an extreme Kerr hole. 

\section{Broad-line-free sources}

Some objects show no evidence for a broad line. Good examples from
long XMM-Newton exposures are Akn\,120, which has no warm absorber
(Vaughan et al. 2004), and the broad line radio galaxy 3C\,120
(Ballantyne et al. 2004). 

Various possibilities for the lack of any line have been proposed by
the authors of those papers including: a) the central part of the disc
is missing; b) the disc surface is fully ionized (ie the iron is); c)
the coronal emissivity function is flat, which could be due to d) the
primary X-ray sources being elevated well above the disc at say $100
r_{\rm g}$.

There are also intermediate sources where the data are either poor or
there are complex absorption components so that one cannot argue
conclusively that there is a relativistic line present. Some narrow
line components are expected from outflow, warm absorbers and distant
matter in the source. One common approach in complex cases, which
is not recommended, is to continue adding absorption and emission
components to the spectral model until the reduced $\chi^2$ of the fit
is acceptable, and then claim that model as the solution. Very broad
lines are difficult to establish conclusively unless there is
something such as clear spectral variability indicating that the
power-law is free of Fe-K features, as found for MCG--6-30-15, or for
GX\,339-4 where the complexities of an AGN are not expected.

\section{Generalization of the light-bending model}

Our interpretation of the spectral behaviour of \mcg\ and some other
sources means that we are observing the effects of very strong
gravitational light bending within a few gravitational radii of a
rapidly spinning black hole. The short term (10--300~ks) behaviour is
explained, without large intrinsic luminosity variability, through
small variations in the position of the emitting region in a region
where spacetime is strongly curved. 

This implies that some of the rapid variability is due to changes in
the source position. Now BHC in the (intermediate) high/soft state
have high frequency breaks at higher frequency, for the same source,
than when in the low/hard state (cf. Cyg X-1, Uttley \& McHardy 2004).
This additional variability when in the soft state is identified with
relativistic light-bending effects on the power-law continuum.

This picture suggests a possible generalization of the light-bending
model to unify the AGN and BHC in their different states.  Note the
work of Fender et al. (2004) which emphasises that jetted emission
occurs commonly in the hard state of BHC. The key parameter may be the
{\sl height} of the main coronal activity above the black hole. Assume
that much of the power of the inner disc passes into the corona
(Merloni \& Fabian 2002) and that the coronal activity is magnetically
focussed close to the central axis. Then at low Eddington ratio the
coronal height is large (say $100r_{\rm g}$ or more), the corona is
radiatively inefficient and most of the energy passes into an outflow;
basically the power flows into a jet. Reflection is then appropriate for
Euclidean geometry and a flat disc and there is only modest broadening to the
lines. If the X-ray emission from the (relativistic) jet dominates
then X-ray reflection is small (see e.g. Beloborodov 1999). The high
frequency break to the power spectrum is low ($\sim 0.001c/r_{\rm
g}$). 

When the Eddington fraction rises above say ten per cent, the height
of the activity drops below $\sim 20 r_{\rm g}$, the corona is more
radiatively efficient and more high frequency variability occurs due
to light bending and the turnover of the power spectrum rises above
$0.01c/r_{\rm g}$. The X-ray spectrum is dominated at low heights by
reflection, including reflection-boosted thermal disk emission, and a
broad iron line is seen. Any jet is weak. 

The objects with the highest spin and highest accretion rate give the
most extreme behaviour. Observations suggest that these include NLS1
and some very high state, and intermediate state, BHC. Some
broad-line-free sources do not however fit this model, so more work is
required.

\section{Summary}

A relativistically-broadened iron line is unambiguous in the spectra
and behaviour of a few objects. The strength and breadth of reflection
features is strong evidence for gravitational light bending and
redshifts from a few $r_{\rm g}$. They indicate that a dense disc
extends close to the black hole, which must therefore be rapidly
spinning ($a/m>0.8$).  

The potential for understanding the accretion flow close to a black
hole is enormous. Current observations are at the limit of
XMM-Newton's powers, which nevertheless has enabled a breakthrough in
understanding the spectral behaviour of \mcg\ and similar objects.
Similarities in the spectral and timing properties of AGN and BHC is
enabling further progress to be made. Studies in the near future with
ASTRO-E2 followed by XEUS and Constellation-X in the next decade will
continue to open up the immediate environment of accreting black
holes, within just a few gravitational radii, to detailed study.

\acknowledgements
Thanks to G. Miniutti, S. Vaughan, K. Iwasawa and J. Miller for
collaboration and many discussions. The generalization of the light
bending modelling was developed with them and A. Merloni. The Royal
Society is thanked for support. 
\theendnotes

\end{article}
\end{document}